 \documentclass[reprint,amsmath,amssymb,aps,pre,showkeys]{revtex4-2} 

\usepackage{color} 
\usepackage{comment} 
\usepackage{epsfig}  
\usepackage{graphicx}
\usepackage{dcolumn}
\usepackage{bm}

\def \ccomma{\raise 2pt\hbox{,}\ } 
\def \D {\hbox{d}}

\def \Re  {\mathop{\rm Re}\nolimits}
\def \Im  {\mathop{\rm Im}\nolimits}

\def \mod#1{\vert #1 \vert}

 \def   \sectio#1    {\section{#1}}
 \def\subsectio#1 {\subsection{#1}}

\def\GLA{A}
\def \csi{\kappa_{\rm i}} 
\def \csr{\kappa_{\rm r}} 
\def \Cthree{K_1} 
\def \Cfour {K_2} 
\def \Mshift{M_0} 
\def \ersurei{\lambda}

\begin{document}

\preprint{Borelli preprint number}

\title{New solutions to the complex Ginzburg-Landau equations}

\author{Robert Conte}
 \homepage{E-mail Robert.Conte@cea.fr\\ ORCID https://orcid.org/0000-0002-1840-5095}
\affiliation{
 Universit\'e Paris-Saclay, ENS Paris-Saclay, CNRS, Centre Borelli, F-91190 Gif-sur-Yvette, France\\
}%
\affiliation{
 Department of Mathematics, The University of Hong Kong, Pokfulam Road, Hong Kong
}%

\author{Micheline Musette}
 \homepage{E-mail Micheline.Musette@gmail.com}
\affiliation{
 Dienst Theoretische Natuurkunde, Vrije Universiteit Brussel, Pleinlaan 2, B–1050 Brussels, Belgium
}%

\author{Tuen Wai Ng}
 \homepage{E-mail NTW@maths.hku.hk\\ ORCID https://orcid.org/0000-0002-3985-5132}
\affiliation{
 Department of Mathematics, The University of Hong Kong, Pokfulam Road, Hong Kong
}%

\author{Chengfa Wu}
 \homepage{CFWu@szu.edu.cn\\ ORCID https://orcid.org/0000-0003-1697-4654}
\affiliation{
 Institute for Advanced Study, Shenzhen University, Shenzhen, PR China
}%

\date{July 03, 2022; revised August 26, 2022, accepted August 31, 2022}
\begin{abstract} 
The various r\'egimes observed in the one-dimensional complex Ginzburg-Landau equation
result from the interaction of a very small number
of elementary patterns such as pulses, fronts, shocks, holes, sinks. 
We provide here three exact such patterns
observed in numerical {\color{red} calculations} but never found analytically.
One is a quintic case localized homoclinic defect, observed by Popp et alii,
the two others are bound states of two quintic dark solitons,
observed by Afanasyev et alii.
\end{abstract}

															
\maketitle

PhySH (Physics Subject Headings):

1. Traveling waves (Primary)|Coherent structures|Pattern formation
2. Optics and lasers-Nonlinear optics-Optical solitons;
3. Nonlinear waves-Solitons;
4. Plasma waves-Solitons;
5. Techniques-Theoretical and Computational Techniques-Field and string theory models and techniques-Classical solutions in field theory-Solitons.




\sectio{Introduction}
\label{sectionIntroPRL}

Slowly varying amplitudes of numerous physical phenomena evolve in time
according to the ubiquitous one-dimensional complex Ginzburg-Landau equation (CGL) equation 
\begin{eqnarray}
& &  {\hskip -18.0 truemm}
i \GLA_t +p \GLA_{xx} +q \mod{\GLA}^2 \GLA +r \mod{\GLA}^4 \GLA -i \gamma \GLA =0,\
\label{eqCGL35}
\end{eqnarray}
in which the constants $p,q,r$ are complex and $\gamma$ is real.

These phenomena include
pattern formation, 
spatio-temporal intermittency, 
superconductivity, 
nonlinear optics, 
Bose-Einstein condensation, 
etc,
see the reviews \cite{AK2002,vS2003}. 

In the cubic case ($r=0$),
it describes for instance the formation of patterns near a Hopf bifurcation
($\gamma$ being the distance from criticality)
to an oscillary state.
The r\'egimes are extremely rich
and, depending on the parameters,
they range from chaotic (turbulent) to regular (laminar),
see the phase diagrams in the plane $(\Re(p)/\Im(p),\Re(q)/\Im(q))$
\cite[Fig.~1]{Chate1994} 
\cite[Fig.~1a]{vanHecke}.
The observed patterns have been classified \cite[Fig.~1]{vSH1992} according to 
both their 
homoclinic 
(equal values of $\lim_{x \to - \infty} \mod{A}$ and  
$\lim_{x \to + \infty} \mod{A}$) 
or heteroclinic (unequal values) nature,
and their topology:
pulses, fronts, shocks, holes, sinks.
For instance, the holes (characterized by the existence of a minimum of $\mod{A}$) 
can be 
either heteroclinic (like the analytic solution of Bekki and Nozaki \cite{BN1985}), 
      or homoclinic (as displayed by numerical simulations of van Hecke \cite{vanHecke}).
Of particular interest is the case when $\mod{A}$ can vanish;
since the phase $\arg{A}$ is then undefined, it can undergo discontinuities,
a feature which creates topological defects.
This ``defect-mediated turbulence'' 
\cite{CGL1989} 
\cite{vHH2001} 
is a major mechanism 
\cite{Shraiman-et-al1992} 
of transition to a turbulent state,
in addition to the mechanism of phase turbulence.
			
The situation is similar in the complex quintic case (CGL5, $r/p$ not real) \cite{vSH1992},
and the importance of these coherent structures
is their role of separators between different r\'egimes,
cf.~\cite[Figs.~5,6]{vSH1992}.
All these coherent structures are indeed observed both in physical phenomena
and in numerical simulations \cite{Chate1994} \cite{vanHecke}.

In Taylor-Couette flows between rotating or counter-rotating cylinders \cite{Latrache_etal2016},
when a parameter varies,
one first observes the expected Benjamin-Feir instability,
followed by the occurence of spatio-temporal defects
(i.e.~a vanishing of $\mod{A}$ which allows a discontinuity in the phase of $A$)
and a large variation of the amplitude.
A similar behaviour is also observed in Rayleigh-B\'enard convection
(a fluid between two conducting plates, heated from below)
\cite{MannevilleBook}.

In nonlinear optics
where $t$ is a coordinate along the fiber
and $x$ a transverse coordinate,
the goal is to carefully select the initial signal
(for instance one of the coherent structures)
and to tune the parameters of the CGL equation in order to
minimize the attenuation during the propagation of this signal.
Various kinds of bright or dark solitons,
or more general ``dissipative'' solitons can be used for this purpose
\cite{Kivshar-Agrawal-Book} 
\cite{Moloney-Newell-Book} 
\cite{AABook2008}. 

There also exist other optical devices
described by a system 
\cite[Eqs.~(1)--(3)]{2006Hachair-et-al.VCSEL}
involving a CGL-like equation for the electric field.
In these 
broad area vertical cavity surface-emitting lasers (VCSELs)
with a single longitudinal mode,
two transverse coordinates (among them our $x$) are necessary
to describe the broad area,
and one observes for instance various patterns 
similar to those in spatiotemporal intermittency
\cite{2017Coulibaly-et-al.VCSEL}
\cite{2022Rimoldi-et-al.VCSEL}.

{}From a more general point of view,
there is a strong evidence \cite{BZvHBT}
that coherent structures govern the transition between different scenarii of chaos.

Of special interest to our purpose is the fact that
some of the observed patterns have been represented by exact, closed form,
analytic expressions, 
such as the CGL5 heteroclinic hole \cite{BN1985},
a traveling wave with arbitrary velocity.
Conversely, it is reasonable to believe 
that every elementary observed pattern
could be associated to some analytic solution, to be found,
this is the motivation of the present work.

In this Letter,
we consider the most challenging situation,
i.e.~the so-called ``complex cases'' in which 
$r/p$ is not real
and, if $r$ vanishes,
$q/p$ is not real,
respectively denoted CGL5 and CGL3.
This excludes the nonlinear Schr\"odinger (NLS) limit ($p,q$ real, $\gamma$ zero),
which does not display any chaotic r\'egime.
Our interest is to look for exact traveling wave solutions
($c$ and $\omega$ real),
\begin{eqnarray}
& & {\hskip -15.0 truemm}
\GLA =\sqrt{M(\xi)} e^{ i(\displaystyle{-\omega t + \varphi(\xi)})},\
\xi=x-ct.
\label{eqCGL35ReducMphi}
\end{eqnarray}
Such traveling waves are characterized by a third order nonlinear
ordinary differential equation (ODE), see (\ref{eqCGL35Order3}) hereafter.

The current list of exact traveling waves is very short,
it comprises only six solutions \cite{CM2005}:
for CGL3, 
a homoclinic pulse \cite{HS1972},
a heteroclinic front \cite{NB1984},
a heteroclinic source/hole \cite{BN1985} \cite[Fig.~5]{Lega2001};
for CGL5,
a homoclinic pulse \cite{vSH1992},
a heteroclinic front \cite{vSH1992},
a homoclinic source/sink \cite{Moores,MCC1994}. 
Other elementary patterns exist, but they have only been observed experimentally,
an important one being a CGL3 homoclinic hole \cite{vanHecke}
\cite[Fig.~1b]{Uchiyama-Konno-2014}.

A lot of effort has been devoted to the search for exact solutions of CGL3/5,
by essentially three methods, which we now outline.
A slight modification of the method of Hirota allowed
Bekki and Nozaki \cite{NB1984} \cite{BN1985} to uncover two of the three above mentioned
traveling waves of CGL3.
By making heuristic assumptions 
among the components of a three-dimensional dynamical system equivalent to (\ref{eqCGL35Order3}),
van Saarloos and Hohenberg \cite[\S 3.3]{vSH1992}
succeeded to obtain two remarkable solutions of CGL5: the pulse and the front solutions.
In the third method, building on previous work \cite{CM1993}, 
Marcq \textit{et alii} \cite{MCC1994} 
made an assumption for the complex amplitude $A$
matching the structure of singularities \cite{CMBook2}
and thus found the full CGL5 source/sink.

In the present work,
by combining two mathematical methods,
we present new traveling waves, outlined in \cite{CMBook2},
and moreover we prove that these are the only ones 
whose square modulus $M(\xi)$
admits only poles as singularities in the complex plane of $\xi$
(in short, is \textit{meromorphic} on $\mathbb{C}$).
Each such meromorphic solution is characterized by a first order nonlinear ODE for $M(\xi)$.

Only three of these new solutions are bounded,
they are all homoclinic,
decrease exponentially fast to a constant value at infinity and only exist for CGL5.
One represents a topological defect,
previously observed experimentally 
\cite[Fig.~3b]{PSAWK1993} \cite[Fig.~4]{PSAK1995}
for a set of parameters compatible with ours.
The two other traveling waves are bound states made of two CGL5 dark solitons;
while in CGL3 the presence of sources inhibits the formation 
of bound states of dark solitons \cite{ACM1998},
such bound states have been numerically observed in CGL5
by Afanasjev \textit{et alii} \cite[Fig.~4]{ACM1998} 
and more recently in 
\cite[Fig.~3d]{Viscarra-Urzagasti-2022}.

The three other solutions,
outlined in \cite{CMBook2} 
(one doubly periodic for CGL3, one doubly periodic and one rational for CGL5) are unbounded,
but,
under a small perturbation
which moves the poles outside the real axis $x-c t$,
their analytic expression becomes bounded,
and the corresponding periodic patterns could be good approximations
to a variety of periodic patterns observed experimentally,
this is currently under investigation.

In the next section,
we simply outline the mathematics involved.
Then, for each bounded solution,
the amplitude $A$ is presented as a product of complex powers
of entire functions.


\sectio{The method}
\label{Section-method}

It arises from a simple remark:
in all presently known exact traveling waves (six, recalled above),
the only singularities of the square modulus $\mod{A}^2=M$ 
in the complex plane of $\xi$ are poles.
Conversely, let us make the \textit{only} assumption that $M(\xi)$ is meromorphic on $\mathbb{C}$.

First, using Nevalinna theory 
\cite{LaineBook},
it has been proven \cite{ConteNgCGL5_ACAP} 
that,
for all values of the CGL parameters $p,q,r$ (complex), $\gamma$ (real)
and of the traveling waves parameters $c,\omega$ (real),
for both CGL3 
and CGL5,     
all solutions $M(\xi)$ meromorphic on $\mathbb{C}$ are elliptic or degenerate elliptic
and therefore obey a nonlinear ODE of first order.
\smallskip

As a consequence, in order to find all meromorphic solutions of both CGL3 and CGL5,
only a finite number of possibilities need to be examined.
This is done 
by two methods.
The first method (Hermite decomposition \cite{Hermite-sum-zeta})
represents $M$ as a finite sum of derivatives of ``simple elements''
admitting only one pole of residue unity
(Weierstrass' $\zeta(\xi)$ \cite{AbramowitzStegun} 
 or its degeneracies $k \coth(k \xi)$ and $1/\xi$),
while the second one (subequation method \cite{MC2003,CM2009})
builds the first order ODE obeyed by $M(\xi)$
then integrates it.
Full details can be found in \cite{CMBook2}.

The modulus $M(\xi)$ obeys a third order ODE \cite{MC2003},
\begin{eqnarray}
& & 
\begin{array}{ll}
\displaystyle{
(G'-2 \csi G)^2 - 4 G M^2  (e_i M^2 + d_i M - g_r)^2=0,
}\\ \displaystyle{
G=\frac{M M''}{2} - \frac{M'^2}{4} 
  -\frac{\csi}{2} M M'  + g_i M^2 + d_r M^3 + e_r M^4, 
}\\ \displaystyle{
\varphi' =\frac{\csr}{2}+\frac{G'-2 \csi G}{2 M^2( g_r - d_i M - e_i M^2)},
}\\ \displaystyle{
}
\end{array}
\label{eqCGL35Order3}
\end{eqnarray}
in which the real parameters $d_r,d_i,e_r,e_i,\csr,\csi,g_r,g_i$ are 
\begin{eqnarray}
\begin{array}{ll}
\displaystyle{
\frac{q}{p}=d_r + i d_i, 
\frac{r}{p}=e_r + i e_i,
\frac{c}{p}=\csr - i \csi,
}\\ \displaystyle{
\frac{\gamma + i \omega}{p}=g_r + i g_i - \frac{1}{2} \csr \csi - \frac{i}{4} \csr^2.
}
\end{array}
\label{eqCGL35Notation}
\end{eqnarray}

This ODE is the key to obtain all meromorphic solutions $M(\xi)$
and, by the quadrature (\ref{eqCGL35Order3})${}_3$,
the complex amplitude $\GLA$.
These solutions occur for specific values of $d_r/d_i$ (CGL3)
and $e_r/e_i$ (CGL5),
which characterize 
the local behaviour of $\GLA e^{i \omega t}$ near a movable singularity $\xi_0$
(i.e.~whose location depends on the initial conditions)
\cite{CT1989,MCC1994},
\begin{eqnarray}
\GLA e^{i \omega t}\sim \left\lbrace
\begin{array}{ll}
\displaystyle{
\hbox{(CGL3) } A_0 (\xi-\xi_0)^{-1+i \alpha},
}\\ \displaystyle{
\hbox{(CGL5) } A_0 (\xi-\xi_0)^{-1/2+i \alpha},
}
\end{array}
\alpha \hbox{ real}.
\right.
\label{eqalpha}
\end{eqnarray}

\section{The new analytic patterns}
\label{sectionNewBounded}

\subsection{CGL5, localized homoclinic defect} 

For the parameters
\begin{eqnarray}
& & 
\begin{array}{ll}
\displaystyle{
\csi=0,\
 \frac{e_r}{e_i}=\frac{3}{2},
 \frac{d_r}{d_i}=\frac{29}{15},
 \frac{g_r}{g_i}=-\frac{12}{35},
      g_i=\frac{7 d_i^2}{12 e_i}\ccomma
}
\end{array}
\label{eqCGL5Trigo4DefectSubeqParam}
\end{eqnarray}
there exists a first order subequation 
\begin{eqnarray}
& & 
\begin{array}{ll}
\displaystyle{
\left({M'}^2 + e_i M \left( M + \frac{2 d_i}{3 e_i}\right) P_2 \right)^2
 -\frac{4}{3} e_i^2 M^2 P_2^3=0,
}\\ \displaystyle{
P_2=M^2 + \frac{6 d_i}{5 e_i} M + \frac{d_i^2}{3 e_i^2}\ccomma
}
\end{array}
\label{eqCGL5Trigo4DefectSubeq}
\end{eqnarray}
in which a rescaling of $M$ and $\xi$ leaves no arbitrariness.
If $p$ is not real, 
as usually assumed so as to be far away from the integrable 
nonlinear Schr\"odinger situation,
this pattern is stationary ($c=0$),
otherwise it is moving with an arbitrary velocity $c$.

In the original notation $(p,q,r,\gamma,c,\omega)$,
the parameters obey the constraints
\begin{eqnarray}
& & 
\begin{array}{ll}
\displaystyle{
c p_i=0,
 \frac{p_r r_r + p_i r_i }{p_r r_i - p_i r_r}=\frac{3}{2}\ccomma
 \frac{p_r q_r + p_i q_i }{p_r q_i - p_i q_r}=\frac{29}{15}\ccomma
}\\ \displaystyle{
\gamma=\frac{(p_r q_i - p_i q_r)^2}{\mod{p}^2 (p_r r_i - p_i r_r)}
 \left(-\frac{7}{12} p_i - \frac{1}{5} p_r\right),
}\\ \displaystyle{
\omega=\frac{(p_r q_i - p_i q_r)^2}{\mod{p}^2 (p_r r_i - p_i r_r)}
 \left(\frac{7}{12} p_r - \frac{1}{5} p_i\right)
 - \frac{p_r^3}{4 \mod{p}^4}c^2.
}
\end{array}
\label{eqCGL5Trigo4DefectSubeqParamOriginal}
\end{eqnarray}

The solution of this subequation 
(the invariance of (\ref{eqCGL35Order3}) by translation
allows us to set $\xi_0=0$),
\begin{eqnarray}
& & 
\begin{array}{ll}
\displaystyle{
M=- 20 \frac{d_i}{e_i} 
 \frac{       \coth^2 \displaystyle\frac{k \xi}{2} 
        \left(\coth^2 \displaystyle\frac{k \xi}{2}-1 \right)}
      {\left(5\coth^2 \displaystyle\frac{k \xi}{2}-3 \right)^2 -12}\ccomma
}\\ \displaystyle{
k^2=\frac{d_i^2}{15 e_i}
=\frac{(p_r q_i - p_i q_r)^2}{15 (p_r r_i - p_i r_r)\mod{p}^2}\ccomma}
\end{array}
\end{eqnarray}
displays four simple poles $\pm \xi_A,\pm\xi_B$,
\begin{eqnarray}
& & {\hskip -7.0 truemm}
\begin{array}{ll}
\displaystyle{
\coth\frac{k\xi_A}{2}= \frac{\sqrt{10\sqrt{3}+15}}{5}, 
\coth\frac{k\xi_B}{2}= \frac{\sqrt{10\sqrt{3}-15}}{5} i,
}
\end{array}
\end{eqnarray}
and 
a shift of $k \xi/2$ by one half-period $i \pi/2$ 
(equivalent to permuting $\cosh$ and $\sinh$)
makes $M$ bounded for $e_i=\Im(r/p)>0$.
In order to deduce the complex amplitude, 
one computes 
the logarithmic derivative $\D \log (A e^{i \omega t})/ \D \xi$
with $A$ defined by (\ref{eqCGL35ReducMphi}).
By virtue of (\ref{eqCGL35Order3})${}_3$,
this is a rational function of $\coth (k \xi/2)$,
whose Hermite decomposition \cite{Hermite-sum-zeta}
is a finite sum of shifted $\coth$.
Then its logarithmic primitive yields  
the complex amplitude $A$ as the product of powers of
five {\color{red} $\sinh$} functions,
\begin{eqnarray}
& & 
\begin{array}{ll}
\displaystyle{
\frac{A}{A_0}=e^{\displaystyle -i \omega t + i \frac{\csr}{2} \xi} 
\sinh \frac{k \xi}{2}
}\\ \displaystyle{ \phantom{12}
\left(\sinh \frac{k(\xi-\xi_A)}{2}\sinh \frac{k(\xi+\xi_A)}{2}\right)
  ^{\displaystyle \frac{-1+(3+2\sqrt{3})i}{2}}
}\\ \displaystyle{ \phantom{12}
\left(\sinh \frac{k(\xi-\xi_B)}{2}\sinh \frac{k(\xi+\xi_B)}{2}\right)
  ^{\displaystyle \frac{-1+(3-2\sqrt{3})i}{2}}\cdot
}
\end{array}
\label{eqCGL5Trigo4DefectA}
\end{eqnarray}

\begin{figure}[ht]
\begin{center}
 \includegraphics[scale=0.3]            {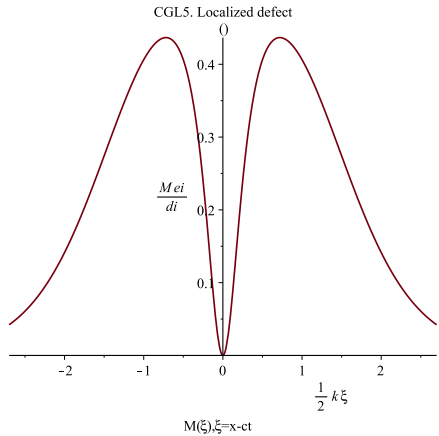}
\end{center}
\caption[CGL5. Homoclinic defect.]
        {CGL5. \hfill\break\noindent Homoclinic defect
				in dimensionless units $M/(d_i/e_i)$ vs.~$k \xi/2$.
				}
\label{FigCGL5-Two-hump-pulse}
\end{figure}

The modulus $M$ displays a unique minimum $M=0$.
This new analytic pattern 
decreases exponentially fast at infinity,
has the topology of a double pulse 
(Fig.~\ref{FigCGL5-Two-hump-pulse}) and
is the first exact representation of a defect in CGL.
The study of its stability under small perturbations has not been performed in this
Letter, this will be investigated later.

Although defect-mediated turbulence is mainly observed in two-dimensional CGL3 
where this is a major mechanism of turbulence
\cite{LegaThese} \cite{Lega2001}, 
it has also been reported in CGL5 \cite[Fig.~3b]{PSAWK1993} \cite[Fig.~4]{PSAK1995}
\cite[p 278]{Lega2001},
where, for a destabilizing CGL5 term 
(negative $\delta$ in the notation of Ref.~\cite{PSAWK1993})
compatible with the present numerical values (\ref{eqCGL5Trigo4DefectSubeqParam}),
one observes a succession of phase slips (every time $M$ vanishes),
which create hole-shock collisions,
ending in a process of as many annihilations as creations.
Such a process is known as ``hole-mediated turbulence''. 
We must also mention that a pattern topologically identical to the present defect
has been numerically observed in a system of two amplitude equations
coupled quadratically \cite[Fig.~2]{DCL2000}. 

This exact pattern should be quite useful for determining the range of parameters
for which topologically similar patterns are stable or metastable.
Indeed, choosing as initial data an analytic expression 
similar to (\ref{eqCGL5Trigo4DefectA}),
for appropriate sets of parameters of CGL5
should considerably shorten the duration of the transient r\'egime
and accordingly increase the convergence towards a topologically similar pattern.

\subsection{CGL5, bound state of two dark solitons} 

When $e_r/e_i$ is one of the four real roots  of
\begin{eqnarray}
& & 
\begin{array}{ll}
\displaystyle{
1089-81327 \ersurei^2+323512 \ersurei^4+456976 \ersurei^6=0, 
}\\ \displaystyle{ 
\ersurei=\frac{e_r}{e_i},
\ersurei=\pm 0.1192,       
\ersurei=\pm 0.4300, 
}      
\end{array}
\end{eqnarray}
there also exists a four-pole subequation (see Appendix for its coefficients) without any free parameter.
To each of these four values of $\ersurei=e_r/e_i$ correspond two values $\alpha_1, \alpha_2$
of the exponent $\alpha$ defined in (\ref{eqalpha}), whose product is $\alpha_1 \alpha_2=-3/4$,
\begin{eqnarray}
& & 
\begin{array}{ll}
\displaystyle{
\ersurei=\frac{e_r}{e_i}=\frac{\alpha}{2}-\frac{3}{8 \alpha},
}\\ \displaystyle{ 
\ersurei=\pm 0.1192        , \alpha=(\mp 0.7550      , \pm 0.9934      ), 
}\\ \displaystyle{ 
\ersurei=\pm 0.4300        , \alpha=(\mp 0.5369      , \pm 1.397      ).  
}
\end{array}
\end{eqnarray}

\begin{figure}[ht]
\begin{center}
 \includegraphics[scale=0.3]            {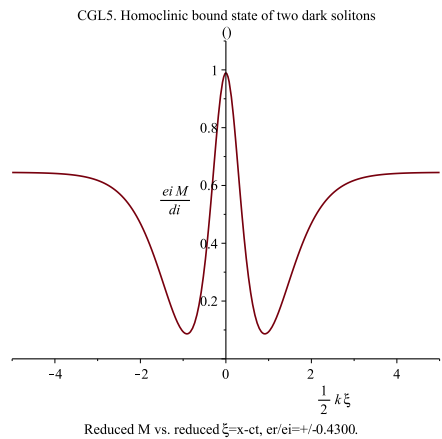}
\end{center}
\caption[CGL5. Homoclinic bound state 1.]
        {\hfill\break\noindent CGL5. The homoclinic bound state 
 in dimensionless units $M/(d_i/e_i)$ vs.~$k \xi/2$, for 
 $\ersurei=\pm 0.4300        $, 
with aspect ratio (1:7.46:11.5). 
	}
\label{FigCGL5-Bound-state-1}
\end{figure}

\begin{figure}[ht]
\begin{center}
 \includegraphics[scale=0.3]            {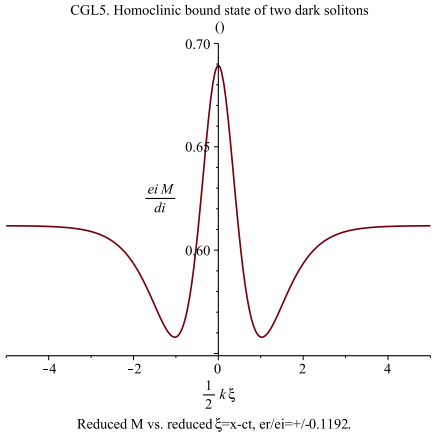}
\end{center}
\caption[CGL5. Homoclinic bound state 2.]
        {\hfill\break\noindent CGL5. The homoclinic bound state 
 in dimensionless units $M/(d_i/e_i)$ vs.~$k \xi/2$, for 
 $\ersurei=\pm 0.1192$, 
}
with aspect ratio (1:1.09:1.23). 
\label{FigCGL5-Bound-state-2}
\end{figure}

The derivation of $M$ then $A$ follows exactly the same logic as for the defect solution.
The complex amplitude $A$ is the product of powers of six $\sinh$ functions
\begin{eqnarray}
& & 
\begin{array}{ll}
\displaystyle{
\frac{A}{A_0}=e^{\displaystyle -i \omega t + i \frac{\csr}{2} \xi} 
}\\ \displaystyle{ \phantom{12}
\sinh \frac{k (\xi-\xi_N)}{2} \sinh \frac{k (\xi+\xi_N)}{2}
}\\ \displaystyle{ \phantom{12}
\left(\sinh \frac{k(\xi-\xi_A)}{2}\sinh \frac{k(\xi+\xi_A)}{2}\right)
  ^{\displaystyle -\frac{1}{2} + i \alpha_1}
}\\ \displaystyle{ \phantom{12}
\left(\sinh \frac{k(\xi-\xi_B)}{2}\sinh \frac{k(\xi+\xi_B)}{2}\right)
  ^{\displaystyle -\frac{1}{2} + i \alpha_2}.
}
\end{array}
\end{eqnarray}
The zeroes ($\pm \xi_A,\pm \xi_B$) have their squares real and are the poles of $M$,
the other zeroes ($\pm \xi_N$) and their complex conjugates $\pm \overline{\xi_N}$
are the four zeroes of $M$,
\begin{eqnarray}
& & {\hskip -5.0 truemm}
\begin{array}{ll}
\displaystyle{
M
= \Mshift + \frac{d_i}{e_i}
 \frac{(\Cthree \coth^2 \displaystyle\frac{k \xi}{2} + \Cfour) (\coth^2 \displaystyle\frac{k \xi}{2}-1)}
      {
			\coth^4 \displaystyle \frac{k \xi}{2} +D_1 \coth^2 \displaystyle \frac{k \xi}{2} +D_0	
			}\cdot  
}
\label{eqM0coth}
\end{array}
\end{eqnarray}
This defines two similar-looking (but different)
homoclinic patterns in the shape of a double well,
whose aspect ratios
($\hbox{min}(M)$:$\lim_{\xi \to \pm \infty} M$:$\hbox{max}(M)$)
are 
(1:7.46:11.5) (Fig.~\ref{FigCGL5-Bound-state-1}) and
(1:1.09:1.23) (Fig.~\ref{FigCGL5-Bound-state-2}).

Like for the defect pattern,
these two patterns are stationary if $p$ is not real
and they move with an arbitrary velocity if $p$ is real.
They compare, at least qualitatively, quite well with the bound state of two CGL5 dark solitons,
as reported apparently for the first time in \cite[Fig.~4]{ACM1998}.


\sectio{Conclusion}

On the numerical side, 
these exact patterns can be used as building blocks to study the interaction
of defects and dark solitons with various other patterns.
On the analytic side,
more exact traveling waves (with less constraints on $p$, $q$, $r$, $\gamma$, $c$, $\omega$)
would necessarily be nonmeromorphic,
this question will be addressed in future work.

\sectio{Acknowledgments}

The authors warmly thank Joceline Lega for sharing her expertise.
The first two authors thank CIRM, Marseille (grant 2311) 
and IHES, Bures-sur-Yvette for their hospitality.
The third author was partially supported by the RGC grant 17307420. 
The last author was supported by the National Natural Science Foundation of China (grant 11701382).

\vfill\eject

\appendix{\textbf{APPENDIX}. Details on CGL5 two homoclinic bound states}


The subequation has the structure
\begin{eqnarray}
& & 
\begin{array}{ll}
\displaystyle{
\left({M'}^2 + c_6 P_{2a}(M) P_{2b}(M)\right)^2-c_7 P_1(M)^2 P_{2a}(M)^3=0,
}\\ \displaystyle{
\csi=0,
P_1   (M)=M         + c_1,
}\\ \displaystyle{
P_{2a}(M)=M^2+c_2 M + c_3,
P_{2b}(M)=M^2+c_4 M + c_5,
}
\end{array}
\label{eqCGL5Trigo4wSubeq}
\label{eqCGL5Trigo4a}  
\end{eqnarray}
with $P_1, P_{2a}, P_{2b}$ polynomials and $c_j$ constants.

Let us choose $e_i$ and $d_i$ as scaling parameters. 
The fixed parameters, as well as three movable constants, 
are polynomials of $e_r/e_i$ (denoted $\ersurei$), 
\begin{widetext}
\begin{eqnarray}
& & {\hskip -5.0 truemm}
\begin{array}{ll}
\displaystyle{
d_r=d_i
\ersurei \frac{828038745921+7649070764998 \ersurei^2+9025535790856 \ersurei^4}{1386644084775}\ccomma
}\\ \displaystyle{
g_r=\frac{d_i^2}{e_i} 
 \frac{-58513290148629717+1113015753503375224 \ersurei^2+1243896610551884848 \ersurei^4}{178728931719095040}\ccomma
}\\ \displaystyle{
g_i=\frac{d_i^2}{e_i} \ersurei
 \frac{119473478956925997-1651180178874084664 \ersurei^2-1567990451264571568 \ersurei^4}{1608560385471855360}\ccomma 
}\\ \displaystyle{
k^2=\frac{d_i^2}{e_i} \ersurei
 \frac{470354925826628997+16800138410952093392\ersurei^4+15744055491100758536\ersurei^2}{2010700481839819200}\ccomma
}\\ \displaystyle{
\Mshift=\frac{d_i}{e_i}
\frac{-344373082347+2958053216864 \ersurei^2+3382994698928 \ersurei^4}{493029007920}\ccomma
}\\ \displaystyle{
\frac{\coth_A}{\coth_B}+\frac{\coth_B}{\coth_A}
 = 2 i \sqrt{3} \ersurei \frac{16223643-41722436 \ersurei^2-37472032 \ersurei^4}{6800175}\ccomma 
}
\end{array}
\label{eqCGL5-csi0-algesvalues}
\end{eqnarray}
\end{widetext}
but $\coth_A=\coth(\xi_A/2), \coth_B=\coth(\xi_B/2)$ 
and the four parameters $\Cthree,\Cfour,D_0,D_1$ of (\ref{eqM0coth})
are algebraic functions of $\ersurei$. 
They are
obtained by equating the rational function (\ref{eqM0coth}) 
and the Hermite decomposition (sum of four simple poles),
\begin{eqnarray}
& & {\hskip -10.0 truemm}
\left\lbrace
\begin{array}{ll}
\displaystyle{
\Cthree            =-\frac{k e_i}{d_i} (a_2 \coth_A+b_2 \coth_B),
b_2=\frac{i \sqrt{3}}{e_i a_2}\ccomma
}\\ \displaystyle{
\Cfour=\frac{k e_i}{d_i} (a_2/\coth_A+b_2/\coth_B) (\coth_A\coth_B)^2,
}\\ \displaystyle{
D_0=\coth_A^2\coth_B^2,
}\\ \displaystyle{
D_1=-(\coth_A^2+\coth_B^2).
}
\end{array}
\right.
\label{eqCGL5-csi0-solCD}
\label{eqCGL5TrigoCjCj}
\end{eqnarray}
All numerical values characterizing the two bound state solutions 
are listed in Table \ref{TableCGL5csi0-wells-num}.

\tabcolsep=0.5truemm

\begin{table}[ht] 
\caption[CGL5 $\csi=0$. Homoclinic bound states.]{
         CGL5 $\csi=0$. Numerical values of the two homoclinic bound states.	
         $e_i$ and $d_i$ are arbitrary real (scaling).
}
\vspace{0.2truecm}
\begin{center}
\begin{tabular}{| l | l | l | l }
\hline 
Variable          & 1 & 2  &   
\\ \hline \hline 
%
%
$e_r/e_i=\ersurei $& $\pm 0.4300              $ &$\pm 0.1192          $& \\ \hline
$d_r/d_i          $& $    0.7910              $ &$    0.08068         $& \\ \hline
$g_r e_i/d_i^2    $& $    1.062               $ &$-0.2375             $& \\ \hline
$g_i e_i/d_i^2    $& $\mp 0.06400             $ &$\pm 0.007092        $& \\ \hline
$\alpha_1         $& $\mp 0.5369              $ &$\mp 0.7550          $& \\ \hline
$\alpha_2         $& $\pm 1.397               $ &$\pm 0.9934          $& \\ \hline
$\coth(k\xi_A/2)  $& $   -0.6093            i $ &$-0.8861            i$& \\ \hline 
$\coth(k\xi_B/2)  $& $    1.259               $ &$ 1.401              $& \\ \hline 
$\coth(k\xi_N/2)  $& $    0.7331              $ &$ 2.319              $& \\ \hline
$\hbox{ (cont'd)} $& $   -0.1430            i $ &$+2.098             i$& \\ \hline
$ (a_2 e_i/d_i) k $& $    0.9531            i $ &$-0.2499            i$& \\ \hline
$\Mshift e_i/d_i  $& $    0.6454              $ &$-0.6118             $& \\ \hline
$\Cthree          $& $   -2.517               $ &$ 0.6230             $& \\ \hline
$\Cfour           $& $    0.2024              $ &$-0.1193             $& \\ \hline
$ D_1             $& $   -1.215               $ &$-1.178              $& \\ \hline
$ D_0             $& $   -0.5889              $ &$-1.541              $& \\ \hline
\end{tabular}
\end{center}
\label{TableCGL5csi0-wells-num}
\end{table}

\vfill\eject
\end{document}